\documentclass[fleqn,10pt]{wlscirep}
\usepackage[utf8]{inputenc}
\usepackage[T1]{fontenc}

\usepackage[modulo]{lineno} 

\usepackage{floatrow}
\usepackage{siunitx}
\usepackage{adjustbox}
\usepackage{blindtext}
\usepackage{multirow}
\usepackage{makecell}

\usepackage{amsmath}
\usepackage{graphicx}
\usepackage[colorinlistoftodos]{todonotes}
\usepackage[colorlinks=true, allcolors=blue]{hyperref}
\usepackage{xcolor}
\usepackage{xspace}
\usepackage{color}
\usepackage{colortbl}
\usepackage{booktabs} 
\usepackage{breakcites} 
\usepackage{comment}
\usepackage{soul}
\newcolumntype{C}[1]{>{\centering\let\newline\\\arraybackslash\hspace{0pt}}m{#1}}

\title{A Semantic Web Technology Index}

\author[1]{Gongjin Lan}
\author[2,*]{Ting Liu}
\author[2]{Xu Wang}
\author[2]{Xueli Pan}
\author[2]{Zhisheng Huang}
\affil[1]{Southern University of Science and Technology, Department of Computer Science and Engineering, Shenzhen, 518055, China}
\affil[2]{VU University Amsterdam, Department of Computer Science, Amsterdam, 1081 HV, the Netherlands}

\affil[*]{t.liu@vu.nl}

\begin{abstract}

Semantic Web (SW) technology has been widely applied to many domains such as medicine, health care, finance, geology.
At present, researchers mainly rely on their experience and preferences to develop and evaluate the work of SW technology.
Although the general architecture (e.g., Tim Berners-Lee's Semantic Web Layer Cake) of SW technology was proposed many years ago and has been well-known, it still lacks a concrete guideline for standardizing the development of SW technology. 
In this paper, we propose an SW technology index to standardize the development for ensuring that the work of SW technology is designed well and to quantitatively evaluate the quality of the work in SW technology. 
This index consists of 10 criteria that quantify the quality as a score of $0\sim10$. We address each criterion in detail for a clear explanation from three aspects: 1) what is the criterion? 2) why do we consider this criterion and 3) how do the current studies meet this criterion?
Finally, we present 
the validation of this index by providing some examples of how to apply the index to the validation cases. 
We conclude that the index is a useful standard to guide and evaluate the work in SW technology. 

\end{abstract}
\begin{document}

\flushbottom
\maketitle
%
%
\thispagestyle{empty}


\section*{Introduction}

The Semantic Web (SW) technology has been widely applied to many domains such as Medicine, Healthcare, Finance, and Geology ~\cite{motta2006next,d2008toward}. 
Knowledge graph is a popular SW technology.
In 2012, Google first promoted a semantic metadata organizational model described as a “knowledge graph” to improve query result relevancy and its overall search experience. 
Since the development of the Semantic Web, knowledge graphs are often associated with linked open data projects, focusing on the connections between concepts and entities.
In recent years, knowledge graphs have grown increasingly prominent in commercial and research applications. 
The SW technology such as knowledge graphs has become an important method in artificial intelligence.

SW technology is an encoding process whereby meaning is stored separately from data and content, which simulates how people understand language and process information. This enables a computer system to have a human-like understanding and reasoning. 
With Semantic Web technologies, adding, changing and implementing new relationships or interconnecting programs in a different way can be as simple as changing the external model that these programs share. 
By approaching the automatic understanding of meanings, SW technology overcomes the limits of other technologies such as the uninterpretable of deep neural networks and their reliance on big training data. 

Although the general architecture (e.g., Tim Berners-Lee's Semantic Web Layer Cake) of SW technology was proposed many years ago and has been well-known, it still lacks a concrete criterion for standardizing the work of SW technology. Currently, the design of SW technology mainly relies on researchers' experience and preferences ~\cite{zolhavarieh2017issues}.
In such a situation, the problems are 1) the work from different researchers are hard to be normalized and be applied by other users and 2) the same work could meet significantly different judgments from different researchers, i.e., inconsistent assessment.
To mitigate the chaotic and non-standardized development, a comprehensive standard is useful and essential to guide and evaluate the design of the work in SW technology. 
Although there are many studies that focus on the evaluation methods, in our knowledge there are no general criteria that standardize the work in SW technology. 
In this paper, we therefore propose an SW technology index that is extracted and refined from the prevalent criteria in the existing works of SW technology ~\cite{berners1998semantic,berners2001semantic}.
The index 
consists of ten criteria with four branches of data, evaluation, knowledge processing, and accessibility, as shown in \autoref{fig:taxonomy}.


With these criteria in this index, users can design the work of SW technology by following the guideline of these ten criteria and quantify the quality as a score of $0\sim10$.
We suggest that researchers should standardize the design of the work in SW technology by following all ten criteria in the development.
For the evaluation of the existing studies, these studies with medium and good level scores can be considered as references for comparability and reproducibility.
These criteria are considered from user-oriented perspectives in terms of science and technology. 
Finally, we 
apply this index to evaluate the popular work
and analysis the scores of these works. 
In summary, 
in this paper we propose an SW technology index guide and evaluate the work in SW technology for designing well with generality, i.e., by following these ten criteria. 
This index with these ten criteria significantly contributes to the normalization of SW technology, which is suitable as introductory material for beginners to the semantic web, as well as supplementary material for advanced users.

\section{Related work}
\label{sec:related}
There are a lot of studies in SW technology over the last decades. 
Many approaches are proposed to design the work of SW technology with a series of criteria from software engineering ~\cite{hooi2015ontology}.
Although many studies investigate how to assess these works, they focus on the evaluation of SW technology. 
By contrast, it is more meaningful how to guide the work in SW technology to be designed well with a good generality.
However, there is a lack of methods, metrics, and tools for improving the development of ensuring that the work of SW technology are designed well. 
This lack of consistent standards is an obstacle to improve the quality of SW technology development and maintenance ~\cite{neuhaus2014toward}.

There are many articles that presented the survey of evaluation methods in SW technology. 
On the Semantic Web, ontology is a key technology that can describe concepts, relationships between entities, and categories of things.
Brank et al. ~\cite{brank2005survey} reviewed the state-of-the-art SW technology evaluations that assess a given ontology from the point of view of a particular criterion of application, typically in order to determine which of several ontologies would be suitable for a specific purpose. 
These papers ~\cite{almeida2011semantics,kuster2008evaluation} presented the surveys on the current evaluation approaches and proposed their evaluation methods for the work of SW technology. 
Yu et al. ~\cite{yu2007ontology} presented a remarkable study in SW technology for evaluation including current methodologies, criteria and measures, and specifically seek to evaluate ontologies based on categories found in Wikipedia. 
Similarly, Hlomani and Stacey ~\cite{hlomani2014approaches} analyzed the state-of-the-art approaches to ontology evaluation, the metrics and measures. 
Verma ~\cite{verma2016abstract} presented a comprehensive analysis of the approaches, perspectives or dimensions, metrics and other related aspects of ontology evaluation.
These articles focus on the evaluation methods for the work of SW (ontology) technology.

Specifically, there are many approaches that were proposed to evaluate the work of SW technology.
Dellschaft and Staab ~\cite{dellschaft2006perform} presented a taxonomic measure framework for gold standard based evaluation of ontologies, which overcomes the problems that the evaluation of concept hierarchies fail to meet basic criteria. 
Brank et al. ~\cite{brank2006gold} proposed an ontology evaluation approach based on comparing an ontology to a gold standard ontology, assuming that both ontologies are constructed over the same set of instances.
Aruna et al. ~\cite{aruna2011survey} proposed an evaluation framework for properties of ontologies and technologies for developing and deploying ontologies.
A famous online tool, Oops! (ontology pitfall scanner!) was proposed for ontology evaluation in ~\cite{poveda2014oops}.
Raad and Cruz ~\cite{raad2015survey} addressed how to find an efficient ontology evaluation method based on the existing ontology evaluation techniques, and presented their advantages and drawbacks.
Similarly, Gao et al. ~\cite{gao2019efficient} proposed an efficient sampling and evaluation framework, which aims to provide quality accuracy evaluation with strong statistical guarantee while minimizing human efforts.
Without exception, these studies focus on the evaluation of the work in SW technology. 
By contrast, we consider the criteria to guide the work of SW technology for designed well with a good generality. 

Finally, there are some studies that provided guidelines with criteria in SW technology. 
However, these guidelines still focus on the evaluation methods in SW technology.
Gangemi et al. ~\cite{gangemi2005theoretical} present a guideline by following several questions for ontology evaluation.
In ~\cite{hooi2015ontology}, a framework is proposed to guide the choice of suitable criteria for various ontology evaluation levels and evaluation methods.
Bandeira et al. ~\cite{bandeira2016foca} provided a guideline with three main steps: ontology type verification, questions verification and quality verification for ontology evaluation.
Sabou and Fernandez ~\cite{sabou2012ontology} provided methodological guidelines for evaluating stand-alone ontologies as well as ontology networks. 
Although these studies provide guidelines, these guidelines focus on how to evaluate the work of SW technology rather than guide how to design the work of SW technology well. 
However, compared with evaluation, it is more meaningful how to guide the work with criteria in SW technology to be designed well with a good generality. 
\section{Methodology (Ten Criteria)}
\label{sec:index}
\begin{figure}[!ht]
    \centering
    \includegraphics[width=0.65\textwidth]{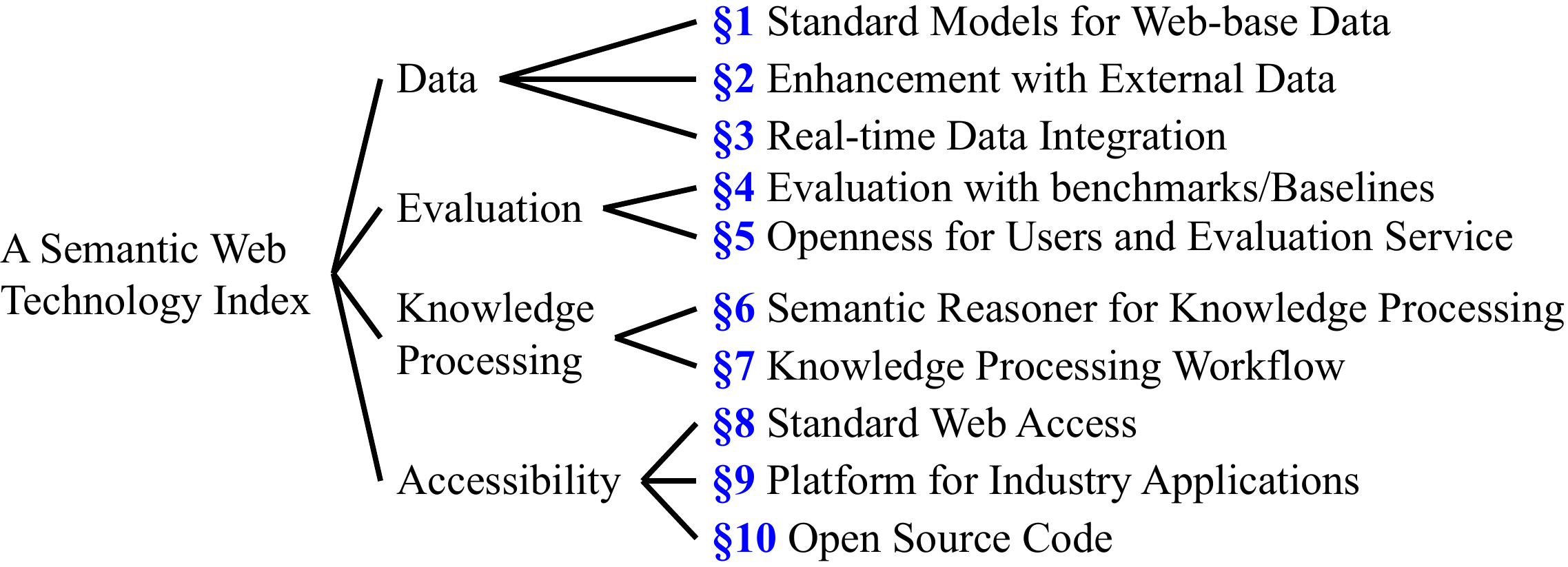}
    \caption{Ten criteria in this Semantic Web technology index.}
    \label{fig:taxonomy}
\end{figure}

We empirically propose the SW technology index with ten criteria that are extracted and refined from the existing works of SW technology ~\cite{berners1998semantic,berners2001semantic}.
Although the index with the ten criteria may not be fully comprehensive, in our knowledge they are solid and complete to guide and evaluate the work of SW technology.
In this section, we address the SW technology index with ten criteria in detail. 
Generally speaking, we can explain an issue by answering the three basic questions of what, why, how.
Therefore, we describe the details of each criterion for a clear explanation in terms of the main three aspects:
\begin{enumerate} [itemsep=-1mm] 
    \item What is the criterion?
    \item Why do we consider the criterion?
    \item How do the current studies meet the criterion?
\end{enumerate}
In particular, the existing studies focus on how to evaluate the work in SW technology that is presented in the fourth criterion, evaluation with benchmarks/baselines in our SW technology index.
We generalize the ten criteria of this index from the four branches of Data, Evaluation, Knowledge Processing, and Accessibility, as listed in \autoref{fig:taxonomy}.

\subsection{Data}
\subsubsection{Standard Models for Web-base Data/Knowledge Representation}
\label{subsec:Model}
The goal of SW technology is to help machines understand data. 
According to the deployment scheme of Five-Star Linked Open Data (\url{https://www.w3.org/2011/gld/wiki/5_Star_Linked_Data}) that is proposed by Tim Berners-Lee, it is essential to create web-base data/knowledge representation by using open standards.
To encode Semantic Web with the data, the well-known standard models of Resource Description Framework (RDF)
and Web Ontology Language (OWL)
are widely applied.

RDF is a popular standard model for structuring web-based data/knowledge. 
It has been used as a general method for conceptual description or modelling of information by using a variety of syntax notations and data serialization formats, particularly used in web-based knowledge management. 
The RDF model manages data by making statements about resources in expressions of three parts: a subject, a predicate, and an object referred to as “triple”, as shown in \autoref{tab:rdf} and \autoref{fig:rdf}. 
Another popular standard model is OWL, which is a family of web-based knowledge representation languages designed to represent rich and complex web-based knowledge about things, groups of things, and relations between things. 
In addition to these two well-known standard models, there are other models for structuring data such as Schema (\url{https://schema.org/})
that is founded by Google, Microsoft, and Yahoo.

\begin{minipage}{\textwidth}
    \begin{minipage}[b]{0.4\textwidth}
    \centering \small
        \begin{tabular}{lll} \toprule
            Subject & Predicate & Object \\ \midrule
            Tim & birthPlace & London \\
            Tim & birthDate & 1955-06-08 \\ \bottomrule
        \end{tabular}
        \captionof{table}{A example of triple statements (RDF).}
        \label{tab:rdf}
    \end{minipage}
\hfill
    \begin{minipage}[b]{0.6\textwidth}
    \centering
    \includegraphics[width=0.6\textwidth]{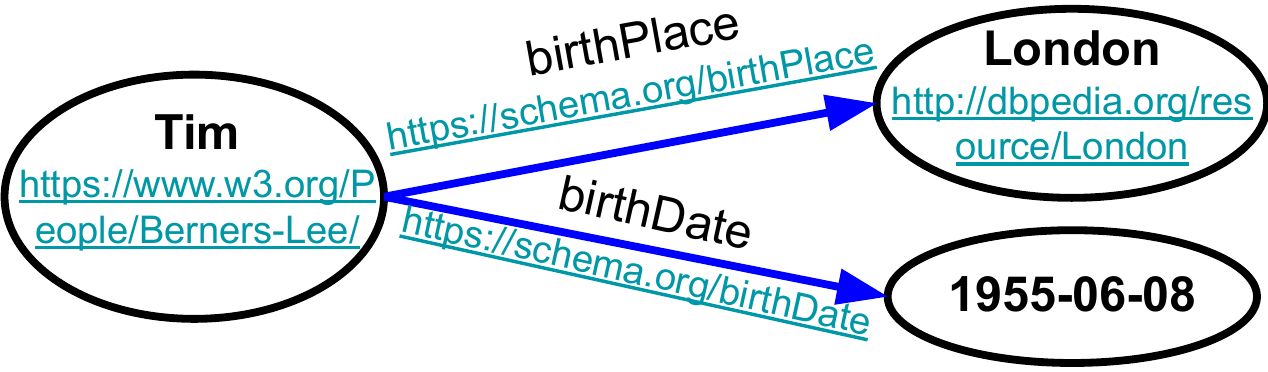}
    \captionof{figure}{The visualization of the RDF in \autoref{tab:rdf}.}
    \label{fig:rdf}
    \end{minipage}
\end{minipage}

These standard models allow structured and semi-structured data to be mixed, exposed, and shared across different applications. 
These technologies formally represent the meaning involved in information. 
For example, ontology can describe concepts, relationships between things, and categories of things. 
These semantics with the data offer significant advantages such as reasoning over data and dealing with heterogeneous data sources. 
These studies can therefore be used by other users easily without the big gap.

Using standard models to manage data has become a consensus for the work in SW technology. 
There are many famous open databases that provide RDF or OWL structured data. For example, Linked Open Data (LOD),
MeSH, 
and Wikidata 
provide RDF structured data. 
The RDF and OWL structured data are available in the DBpedia 
database.  
However, most current databases are stored and updated without semantics, i.e., legacy non-RDF databases (e.g., relational databases). 
For such cases, some approaches have been applied to integrate legacy non-RDF relational databases as virtual RDF graphs, e.g., the well-known Virtuoso ~\cite{erling2009rdf} or D2R ~\cite{bizer2004d2rq} (\url{http://d2rq.org/}) platform. 
Finally, Cryptography is important to ensure and verify that Semantic Web statements are coming from trusted sources. 
In semantic web technology, Cryptography can be generally achieved by the appropriate digital signature of RDF statements. 

\subsubsection{Enhancement with External Data/Knowledge Resources}
\label{subsec:correlation}

There are a lot of overlapping databases that have been reinvented.
We all know that does not reinvent the wheel.
In the second criterion of this index, we therefore suggest that researchers do not spend time to create the data that we can obtain from available databases.
As proposed in the well-known deployment scheme of Five-Star Linked Open Data by Tim Berners-Lee, it is an important criterion that data should be linked to other Linked Open Data.
Therefore, we believe that relevance to existing external databases is an important criterion for the work in SW technologies.

Technically speaking, the correlation of entities in a database with their counterparts in external databases is the key to SW technologies.
In recent years, the volume of Semantic Web data has shown a significant growth trend.
Currently existing Semantic Web databases include not only general Semantic Web-based datasets, such as DBpedia
~\cite{auer2007dbpedia,lehmann2015dbpedia} 
and YAGO 
~\cite{tanon2020yago}, 
but also domain-specific knowledge bases, such as MusicBrainz
and DrugBank.
In SW technologies, the association between Semantic Web and existing popular databases is indispensable.

Semantic Web data aims to describe various entities or concepts.
These entities and concepts are generally structured in databases with unique identifiers.
Thus, the attributes of entities can be enriched by associating the counterparts of entities contained in external data sources.
The connection to external databases makes entities consistent and extends the database, which provides an efficient way to enrich the data/knowledge of the Semantic Web.

\begin{figure}[!ht]
    \centering
    \includegraphics[width=0.45\textwidth]{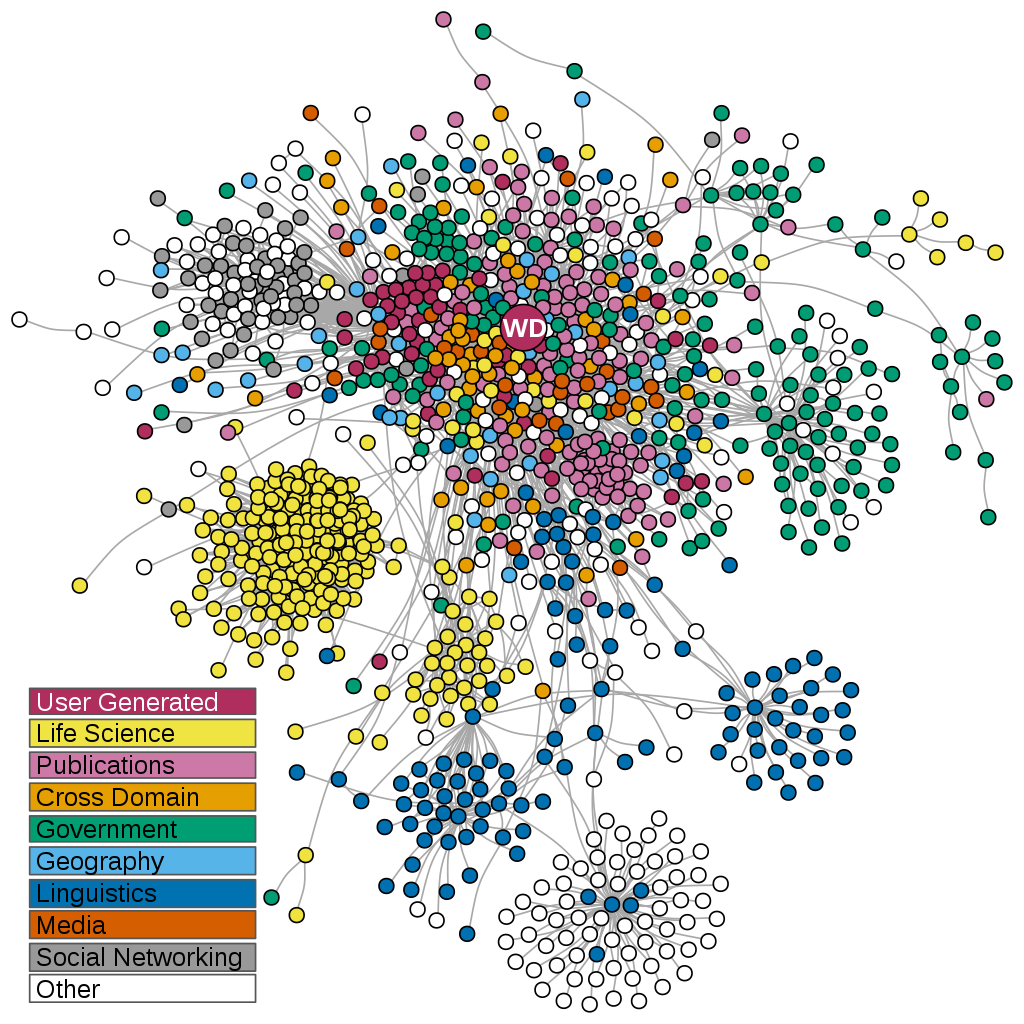}
    \caption{Illustration of Wikidata in the Linked Open Data Cloud ~\cite{vrandevcic2014wikidata}. Databases indicated as circles (with wikidata indicated as ‘WD’), with grey lines linking databases in the network if their data is aligned.}
    \label{fig:wikidata}
\end{figure}

Currently, popular databases have generally established connections to external databases.
For example, DBpedia ~\cite{auer2007dbpedia,lehmann2015dbpedia} and Wikidata ~\cite{vrandevcic2014wikidata} are the well-known large linked open data sets that link entities to their counterparts in external databases.
Wikidata in the linked open data is illustrated in \autoref{fig:wikidata}.
Many researchers associate their data with popular databases when building knowledge graphs.
For example, in ~\cite{liu2020exploring}, the authors enriched their knowledge base by linking to databases UMLS,
MeSH,
and SNOMED CT.
The authors save time for constructing the taxonomy of gut microbiota, since these databases contain taxonomic information on microorganisms.
Therefore, the association with external databases benefits to save time and improve efficiency.
In summary, it is necessary to establish links to external databases.

\subsubsection{Real-time Data Integration}
\label{subsec:itegration}

Data integration is a process dealing with automatic detection, association, correlation, estimation and combination of data from single/multiple sources to produce more consistent, accurate, and useful information than that provided by any individual data source ~\cite{klein2004sensor}.
Data integration for Semantic Web is concerned with the process of building/updating a database.
Real-time data integration for Semantic Web requires high-speed processing during the operation of building/updating.

Although real-time data integration is not requested in the well-known deployment scheme of Five-Star Linked Open Data, it plays a crucial role in updating databases for real applications with real-world data.
For the Semantic Web designed by the real-world data, real-time data integration allows Semantic Web to respond to the change of real-world data.
For instance, DBpedia is a huge database based on Wikipedia data ~\cite{lehmann2015dbpedia} that is edited over a hundred times per minute.
However, DBpedia is usually updated once per month.
DBpedia therefore can be used in real-time by users.
In current studies, there is a lack of Semantic Web that supports real-time data integration.
There are only a few studies that implemented real-time data integration in the building/updating database.
For example, LOD Laundromat
is an open-source toolkit for crawling and cleaning Linked Open Data ~\cite{Wouter2014LOD}.

\subsection{Evaluation}

\subsubsection{Evaluation with Benchmarks/Baselines}
\label{subsec:case}

Evaluation with benchmarks (baselines) is a widely used method in scientific research.
A research design usually involves qualitative and/or quantitative methods ~\cite{josefsson2016good}. 
Evaluation is vital to describe, compare and evaluate a research design from different aspects.
Popular evaluation methods are usually considered for test cases on well-known databases to obtain convincing results.
In general, an evaluation should use multiple cases to compare and elucidate a research design from various aspects, thus to observe whether it meets the requirements and functions correctly.

In scientific research, evaluations in different cases are generally used to reveal errors or defects of used methods.
It is a method to observe if the study evaluated by the cases meets the expected criteria. 
The results from different cases must meet the criteria for establishing trustworthiness in qualitative or quantitative evaluations. 
According to the evaluation with benchmarks, the studies in SW technology from different authors therefore are comparable. 
That is crucial for scientific evaluation and academic exchange.
In SW technology, most studies are conducted by evaluations with benchmarks, as the statistical results (fourth criterion) are shown in \autoref{tab:demon}. 
All of the work except ~\cite{yu2017knowledge} present evaluations with benchmarks. 
Currently, evaluations with benchmarks in SW technology has become a well-known consensus.

\subsubsection{Openness for Users and Evaluation Service}
\label{subsec:openness}
One of the great advantages of Semantic Web technologies is to enable data integration and data reuse. 
To ensure the quality of data, openness for users to access the data and availability for users to access the evaluation services are greatly important in the perspective of semantic collaboration ~\cite{janev2011applicability}.

Differ with open-source, openness for users and evaluation service is concerned with access to use SW technology  and access to evaluate the quality of SW technology with user's data.
Openness for using requires the work of SW technology to provide an interface (e.g., API) for users.
Openness for evaluation allows users to evaluate SW works with other data including user's data.

Openness gives users convenience to use the existing work of SW technology for their work/research.
Openness could provide an interface for users to run experiments on their data with given SW technology.
Openness for evaluation service would bring benefit to the providers of the work of SW technology.
Providers could evaluate the work of SW technology with other user's data.

\subsection{Knowledge Processing}
\subsubsection{Semantic Reasoner for Knowledge Processing}
\label{subsec:Reasoner}

\par
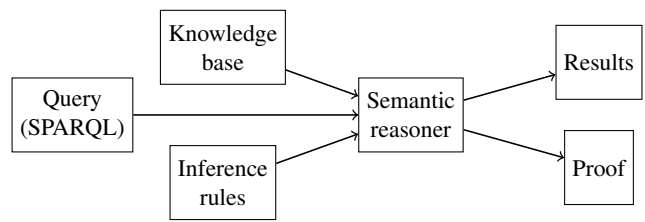
\begin{figure}[!ht] \small
\caption{Illustration of Semantic reasoning process and the various components.}
\label{fig:reasoner} 
{
\begin {tikzpicture}
    \node[draw,align = center,minimum height = 28pt] at (-6.0, 0)  (a) {Query \\ (SPARQL)};
    \node[draw,align = center,minimum height = 28pt] at (-1.5, 0)   (b) {Semantic\\reasoner};
    \node[draw,align = center,minimum height = 28pt] at (-4.0, 0.9)   (c) {Knowledge\\base};
    \node[draw,align = center,minimum height = 28pt] at (-4.0, -0.9)   (d) {Inference\\rules};
    \node[draw,align = center, minimum height = 28pt] at (1.0, 0.7)   (e) {Results};
    \node[draw,align = center, minimum height = 28pt] at (1.0, -0.7)   (f) {Proof};
    \draw[->,semithick] (a) node[above,xshift=1.05cm] {} -- (b);
    \draw[->,semithick] (b) node[above,xshift=1cm] {} -- (e);
    \draw[->,semithick] (b) node[above,xshift=1.55cm] {} -- (f);
    \draw[->,semithick] (c) node[above,xshift=1cm] {} -- (b);
    \draw[->,semithick] (d) node[above,xshift=1cm] {} -- (b);
	\end {tikzpicture}
}
\end{figure}

 Reasoning capability is crucial to many applications developed for the Semantic Web. 
A semantic reasoner is a program that infers logical consequences from a set of facts, assertions, and axioms, which could provide solutions for reasoning tasks such as classification, consistency checking and querying ~\cite{dentler2011comparison,khamparia2017comprehensive}. 
A semantic reasoner provides a plentiful set of mechanisms to generalize an inference engine. 
The inference is normally described by Semantic Web language or description logic language. 
Many reasoners deal with objects by using first-order predicate logic to perform reasoning. 
The inference commonly proceeds by forward and backward chaining. 
In general, a semantic reasoner comprises a knowledge base and an inference engine ~\cite{Chowdhary2020}.
The general semantic reasoning process and the various components are illustrated in \autoref{fig:reasoner}.

Automatic reasoning support is greatly important for Semantic Web services.
On the one hand, it enables correctness checking within ontologies. 
On the other hand, it enables the relation completeness within ontologies, which could improve the quality of the data for a better performance of downstream tasks such as querying and classification.
There are many notable semantic reasoners including CB, CEL, FaCT++.
These reasoners vary significantly from different characteristics. 
It is crucial to design a critical assessment and evaluation to meet different requirements before selecting a reasoner for real-life applications and services. 
Therefore, we consider semantic reasoners as a criterion for SW technology.

\subsubsection{Knowledge Processing Workflow}
\label{subsec:support}

Knowledge processing workflow in designing Semantic Web is a development with multiple processing steps from knowledge to results.
The processing steps generally contain concept/knowledge query, the similarity between knowledge, and knowledge matching.
Designing the work of SW technology by following knowledge processing workflow is very crucial as the two advantages below:
\begin{itemize}[itemsep=-1mm]
    \item the work of SW technology with the knowledge processing workflows is trusty because of its interpretability compared with black-box models.
    \item the potential associations in the works under the knowledge processing workflow can be revealed with the explicit knowledge associations in Semantic Web. 
\end{itemize}

Knowledge query is a popular processing step in the knowledge processing workflow.
The SPARQL query language 
is widely used for knowledge query.
Semantic similarity is widely used in knowledge processing workflow for the work of SW technology.
Many semantic similarity approaches were provided in the last decades ~\cite{wu1994verbs,Resnik1995Resnik,lin1998information}.
Alani et al. ~\cite{Alani2006AKTiveRank} focuses on the investigation of ontology selection and introduces several particularly useful knowledge processing tasks, such as CMM measure (topic coverage), DEM measure (knowledge certain degree decision), and BEM measure (knowledge importance decision).

\subsection{Accessibility}
\subsubsection{Standard Web Access}
\label{subsec:Endpoints}

\par It is crucial that structured data in SW technologies can be accessed from endpoints.
Providing endpoints with standard Semantic Web query services, such as SPARQL endpoints, has become a well-known standard for SW technologies.
As proposed in \autoref{subsec:Model}, structured web-based data/knowledge is generally represented in OWL or RDF.
The primary query languages for OWL ontologies are Semantic Query Augmented Web Rule Language and Description Logic.
SPARQL is a popular query language (protocol) for accessing, retrieving, and manipulating structured data in RDF.
SPARQ is also valid for querying OWL ontologies that are serialized into RDF.

\par Endpoints with standard Semantic Web query language are web services (i.e., HTTP services) that receive and process query protocols and requests.
It provides several benefits in reaching the data.
First, it responds to SPARQL queries without storing the data in the local system.
Second, it integrates HTTP documents thus allowing data interaction.
Third, it can be accessed with other user agents or services, which enables users to create mashups with the data.
Fourth, it brings extensive flexibility to data queries due to the use of HTTP document URLs.
Finally, it supports outputting query scheme files in multiple formats, such as HTML, JSON, CSV, RDF-Turtle, RDF-N-Triples.

\par
Currently, there are various existing well-known databases accessible through SPARQL endpoints, such as Wikidata,
DBpedia,
MeSH, 
and DrugBank.
In addition, the SPARQL query provides application programming interfaces (e.g., 
Yasgui Triply
and Virtuoso SPARQL Query Editor)
for performing federated queries on multiple datasets.
Other graph database platforms or management systems, such as GraphDB and Neo4j, allow users to store, query, and analyze data locally.
It is better to provide a SPARQL endpoint than a huge RDF dump for publishing research reports.
In summary, providing endpoints with standard Semantic Web query languages to reach data is crucial in SW technologies.

\subsubsection{Platform for Industry Applications}
\label{subsec:Platform}

The industrial semantic data processing platform needs to be compliant with semantic data standards, which generally include several features: semantic data storage, semantic query, and basic semantic extension.
For industrial platforms, semantic data storage needs to meet billions of data storage capabilities.
The platform should be applied to the application by semantic query technology (e.g., SPARQL).
In addition, the platform should perform the semantic extension capabilities to implement basic semantic reasoning.

The industrial semantic data processing platform is crucial for SW technology because of its features as below.
1) The storage capacity of massive data allows users to store and process large amounts of data.
2) Semantic query function enables users to query data more effectively, especially massive semantic data.
3) Semantic expansion can be used to infer more potential semantic data, thereby enriching the original data.
To date, there are several popular industrial Semantic Web platforms, such as Oracle 19c,
MarkLogic,
Virtuoso,
Amazon Neptune, 
and Stardog.
These industrial Semantic Web platforms have been used in many works of SW technology.

\subsubsection{Open Source Code}
\label{subsec:open}
In computer science, we often describe programs as being “open source”. Perens et al. ~\cite{perens1999open} proposed 10 criteria to define open source. 
In this paper, their source code should be freely available to users if a program of SW technology is open source. 
The users should be free to download source code, modify it, and distribute their own versions of the program for the desired purpose.

Open source code is important in computer science because it provides a number of benefits to both users and contributors. We summarize two benefits in terms of users and contributors as below.
\begin{enumerate}[itemsep=-1mm]
    \item The studies with open source code are disseminated easily. Users can customize it to fit any purpose. Open source advocates tend to share knowledge and "improve together".
    \item The source code is open to be examined and its defects are also open to be exposed. Any user can discover and fix them. The contributors may not need a long time and cost extra for examination and fixing bugs. 
\end{enumerate}

Open source code is a widely accepted consensus for SW technology. 
There are many studies that open their source code to users. 
For instance, Xiong et al. ~\cite{xiong2017deeppath} and Ding et al. ~\cite{ding2018improving} presented the studies in SW technology and released the source code that can be used easily for other users.
Generally, Github is the popular platform to release the source code. 
On the contrary, although some studies ~\cite{batet2014semantic,wickramarachchi2020evaluation,gu2020ontology} presented semantic-based studies, the source code has not been opened. 
These studies therefore are hard to be applied for users. 
In summary, we consider open source code as an important criterion in our SW technology index.

\section{Validation}
\label{sec:demonstration}
In this paper, we empirically propose an SW technology index with ten criteria.
Although we propose these ten criteria that are extracted and refined from the prevalent criteria in the existing works of SW technology, we still need to validate the rationality of these ten criteria on the existing applications of SW technology.
In this section, we selected and reviewed the existing representative studies in SW technology to explore how do these studies match each criterion for revealing the situation of the work in SW technology.
We apply the index to evaluate the existing studies of SW technology.

To validate the rationality of this index, we select the remarkable and popular studies (e.g., DBpedia, YAGO) in SW technology and evaluate them by these ten criteria of this index with a quantified score of $0 \sim 10$. 
In this paper, we select 16 representative studies in total for the validation. 
For each study, these ten criteria are applied to match the design of these studies.
The score of a study is added by one point for each match.
For example, a study matches the tenth criterion of open source code if the source code of this study is open and free to use and thus its score is added by one point.
As shown in \autoref{tab:demon}, each match (un-match) is donated as "+ (-)".
The sum of matches is the score of a study for the quantitative evaluation, i.e., $0 \sim 10$.
The selected studies are divided into three levels of good ($9 \sim 10$), medium ($6 \sim 8$), and poor ($0 \sim 5$) according to their scores.
Finally, we calculate the rate of studies in three levels separately and the rate of matches of each criterion. 

The results are shown in \autoref{tab:demon}.
We test the selected 16 studies of SW technology with the ten criteria.
Three of them obtain a score of $9 \sim 10$ within the good level, which the rate of the studies in all 16 studies is 18.75\%. 
Seven of the 16 studies obtained medium levels scores (i.e., $6 \sim 8$), that is at a rate of 43.75\%.
Six of the 16 studies fall in the poor levels with $3 \sim 5$ scores, that is in a rate of 37.5\%. 
For the results of each criterion, 13 of the 16 studies match the first, second, and fourth, i.e., a rate of 81.25\%. 
These three criteria are the best matched in the studies.
By contrast, the third and eighth criteria are matched by 8 of the 16 studies, i.e., a rate of 50\%.
The other five criteria are matched by more than 50\% but less than the best of 81.25\%.
These ten criteria in the SW technology index cover the mainstream criteria in these studies.
However, many current studies only applied some of these ten criteria to the design of SW technology and therefore get low scores. 
We conclude that 1) the current situation of the work in SW technology is with casual even chaotic and non-standardized development and 2) the index is a useful standard to guide and evaluate the work of SW technology. 
We will further provide an in-depth discussion of the results in \autoref{sec:discussion}.

\begin{table}[!ht]
    \centering \small
    \renewcommand{\arraystretch}{1.0}
    \setlength\tabcolsep{2pt}
    \begin{tabular}{c l c c c c c c c c c c c c}
        \toprule
        \multirowcell{2}{Studies} & \multirowcell{2}{Authors, year} & \multicolumn{3}{|c|}{Data} & \multicolumn{2}{c|}{Evaluation} & \multicolumn{2}{c|}{Knowledge Process} & \multicolumn{3}{c|}{Accessibility} & \multirowcell{2}{Score} & \multirowcell{2}{Rate} \\ 
        & & \multicolumn{1}{|c}{1} & {2} & \multicolumn{1}{c|}{3} & {4} & \multicolumn{1}{c|}{5} & {6} & \multicolumn{1}{c|}{7} & {8} & {9} & \multicolumn{1}{c|}{10} &  &  \\ \midrule
        \cite{auer2007dbpedia} & Auer et al., 2007 & {\bf +} & {\bf +} & {\bf +} & {\bf +} & {\bf +} & {\bf +}  & {\bf +} & {\bf +} & {\bf +}  & {\bf +} & 10 & \multirowcell{3}{\small 18.75\%}\\
        \cite{tanon2020yago} & Tanon et al., 2020 & {\bf +} & {\bf +}  & {\bf +} & {\bf +} & {\bf +} & {\bf +}  & {\bf +} & {\bf +} &{\bf +} & {\bf +} & 10 &  \\ 
        \cite{liu2020exploring} & Liu et al., 2020 & {\bf +} & {\bf +} & {\bf -} & {\bf +} & {\bf +} & {\bf +} & {\bf +} & {\bf +} & {\bf +} & {\bf +} & 9 & \\  \midrule
        \cite{dhandapani2021question} & Dhandapani et al., 2021 & {\bf +} & {\bf +} & {\bf -} & {\bf +} & {\bf +} & {\bf -} & {\bf +} & {\bf +} & {\bf +} & {\bf +} & 8 & \multirowcell{8}{\small 43.75\%}\\ 
        \cite{bean2017knowledge} & Bean et al., 2017 & {\bf +} & {\bf +} & {\bf -} & {\bf +} & {\bf +} & {\bf +} & {\bf +} & {\bf -} & {\bf +} & {\bf +} & 8 & \\
        \cite{shi2017semantic} & Shi et al., 2017 & {\bf +} & {\bf +} & {\bf +} & {\bf +} & {\bf -} & {\bf +} & \textbf{+} & \textbf{-} & \textbf{+} & {\bf -} & 8 & \\
        \cite{malas2019drug} & Malas et al., 2019 & {\bf +} & {\bf +} & {\bf +} & {\bf +} & {\bf +} & {\bf -} & \textbf{+}& \textbf{-}  & \textbf{+} & {\bf +} & 8 & \\
        \cite{Xu2020Building} & Xu et al., 2020 & {\bf -} & {\bf -} & {\bf +} & {\bf +} & {\bf +} & \textbf{+} &\textbf{+} & \textbf{-} & \textbf{+} & {\bf +} & 7 & \\ 
        \cite{liu2020predicting} & Liu et al., 2020 & {\bf +} & {\bf +} & {\bf -} & {\bf -} & {\bf +} & {\bf -} & {\bf +} & {\bf +} & {\bf -} & {\bf +} & 6 & \\
        \cite{sosa2019literature} & Sosa et al., 2019 & {\bf +} & {\bf +} & {\bf +} & {\bf +} & {\bf +} & \textbf{+} & \textbf{-} & \textbf{-} &  {\bf -} & {\bf -} & 6 & \\ \midrule
        \cite{rotmensch2017learning} & Rotmensch et al., 2017 &  {\bf +} & {\bf +} & {\bf -} & {\bf +} & {\bf -} & {\bf -} & \textbf{+} & \textbf{+} & \textbf{-} & {\bf -} & 5 & \multirowcell{5}{\small 37.5\%} \\ 
        \cite{yu2017knowledge} & Yu et al., 2017 & {\bf +} & {\bf -} & {\bf -} & {\bf -} & {\bf +} & {\bf +} & \textbf{-} & \textbf{+} &\textbf{-}  & {\bf -} & 4 & \\
        \cite{weng2017framework} & Weng et al., 2017 & {\bf +} & {\bf +} & {\bf -} & {\bf +} & {\bf -} & {\bf +} & \textbf{+} & \textbf{-} & \textbf{-} & {\bf -} & 4 & \\
        \cite{chen2020robustly} & Chen et al., 2020 & {\bf -} & {\bf +} & {\bf +} & {\bf +} & {\bf -} & {\bf -} & \textbf{-} & \textbf{-} & \textbf{+} & {\bf -} & 4 & \\
        \cite{hasan2016clinical} & Hasan et al., 2016 &  {\bf -} & {\bf +} & {\bf +} & {\bf +} & {\bf -} & \textbf{-} &\textbf{-} & \textbf{-} & \textbf{-} & {\bf -} & 3 & \\ 
        \cite{rubio2021indoor} & Rubio et al., 2021 & {\bf +} & {\bf -} & {\bf -} & {\bf -} & {\bf -} & {\bf -} & {\bf -} & {\bf +} & {\bf -} & {\bf +} & 3 & \\ \midrule
        \multicolumn{2}{c}{Rate of \bf +} & {\small 81.25\%} & {\small 81.25\%} & {\small 50\%} & {\small 81.25\%} & {\small 62.5\%} & {\small 56.25\%} & {\small 68.75\%} & {\small 50.0\%} & {\small 56.25\%} & {\small 56.25\%} &  & \\
        \bottomrule 
    \end{tabular}
    \caption{The validation of ten criteria in the Semantic Web technology index. All 16 selected representative studies are evaluated with a quantified score of $0 \sim 10$. $+$ ($-$) donates that the study matches (unmatches) the corresponding criterion.}
    \label{tab:demon}
\end{table}

\section{Discussion}
\label{sec:discussion}


Researchers currently develop the works in SW technology based on their experience and preferences. 
Therefore, the design of SW works are generally different.
The studies in \autoref{tab:demon} perform significantly different scores because of their different development. 
The results indicate that the current works of SW technology are casual even chaotic. 
These studies therefore are not easy to be employed by other users and compared each other. 

On one hand, we discuss the positive side of the results.
Most of the 16 selected studies match the ten criteria within the medium and good level scores, that is ten of 16 studies with a rate of 62.5\% (18.75 + 43.75\%).
In particular, Auer et al. ~\cite{auer2007dbpedia} and Tanon et al. ~\cite{tanon2020yago} perform the best score of 10. 
Three (18.75\%) of the 16 studies meet the $9 \sim 10$ criteria, which are with good level scores.
For each criterion, the first (standard models, e.g., RDF and OWL), second criterion (external data), and fourth (evaluation with benchmarks/baselines) are adopted for most selected studies with a rate of 81.25\%. 
Apparently, these three criteria are widely adopted in the current work of SW technology. 
Specifically, RDF and OWL are the well-known standard models that have been widely applied to encode the data in SW technology.
Enhancement with external data is generally adopted to collect data.
The evaluation with benchmarks/baselines is a widely applied approach, in particular scientific research. 
These studies, which are evaluated with benchmarks/baseline, therefore can be assessable by other researchers.

On the other hand, the results indicate the negative side that the current situation of the work in SW technology is with casual even chaotic and non-standardized development.
The 16 studies obtain a wide range of scores $3 \sim 10$ for matching these ten criteria.
Six (37.5\%) of the 16 studies meet only a few criteria that are with the poor level score. 
We notice that both the studies with good and poor level scores consist of the latest work and the early work. 
The two studies with a score of 10 are released in 2007 and 2020. 
The one ~\cite{rubio2021indoor} with a score of 3 is published in 2021. 
Therefore, the design of the studies in SW technology relies on the experience and preferences of different researchers regardless of when the studies are finished. 
The reason can be explained as there are no general criteria to standardize the developments in SW technology up-to-date. 
Some of the selected studies were designed with few criteria, not even using the widely applied criterion such as evaluation with benchmarks/baselines. 
For example, Yu et al. ~\cite{yu2017knowledge} has not opened the source code, not even used the evaluation with benchmarks/baselines. 
For the tenth criterion (open source code), only approximately half of the studies released their source code.
We suggest that the studies should open source code as they can because of the benefits as shown in \autoref{subsec:open}.

For this index, we count the match of each criterion with a binary value if one of these criteria is matched or not within the work of SW technology and sum the number of matches as the total score.
While these ten criteria could be unequal in real applications, we assume that these ten criteria are equally important.
This assumption is mainly based on the following two reasons.
1) We empirically extracted and refined these ten criteria from the existing work of SW technology ~\cite{berners1998semantic,berners2001semantic}, e.g., Tim Berners-Lee's Semantic Web Layer Cake.
However, these ten criteria have not been generalized as a standardized guideline.
To date, these criteria are generally considered equally. 
Therefore, we tend to objectively extract and generalize a scheme of the index with these criteria as this assumption of the same importance rather than redefine a subjective scheme with different degrees for the ten criteria by our experience and preferences.
2) It is likely to fail in objectivity and availability (unavailable if too complex) if we design the different degrees for the ten criteria.
In addition, we consider this issue in the further work that could be updated in the new version. 
Finally, we suggest that researchers should follow all ten criteria in the development to standardize the work in SW technology.
For the evaluation of the existing studies, we suggest that these studies with medium and good level scores can be considered as references for comparability and reproducibility.
The studies with poor level scores are generally hard to be reference for comparability and reproducibility.

Last but not least, the index with the ten criteria may not be fully comprehensive. 
Cryptography is important in semantic web technology, and is considered in the Semantic Web stack/layer cake. 
However, top layers in the Semantic Web stack/layer cake contains technologies, such as Cryptography and Trust, that are not yet standardized or contain just ideas that should be implemented in order to realize Semantic Web ~\cite{ardizzone2012science}.
Moreover, we checked that none of the 16 selected studies in \autoref{tab:demon} considered Cryptography in their developments.
In this paper, we address Cryptography as a part of \autoref{subsec:Model}.
These criteria in this index could be updated in future over the development of SW technology.


We conclude that the current works in SW technology are with casual even chaotic and non-standardized development and hardly to be used for users. 
Such chaotic phenomena could be caused by that there are no generic criteria to standardize the work of SW technology to date. 
The index with ten criteria in this paper provides a feasible solution to mitigate this chaotic situation by providing a standardized guideline for the work in SW technology. 
Therefore, this SW technology index is meaningful for guiding and evaluating the work in SW technology for the normalization of SW technology.

\section{Conclusion}
\label{sec:conclusion}
In this paper, we present an SW technology index that consists of ten criteria. 
On one hand, this index can be used to guide and standardize the design of ensuring that the studies in SW technology are designed well. 
On the other hand, this index quantifies the studies of SW technology as a score of 0 $\sim$ 10. 
we suggest that researchers should standardize the work of SW technology by following all ten criteria for comparability and reproducibility.
For the evaluation on the existing studies, these studies with medium ($6 \sim 8$) and good ($9 \sim 10$) level scores can be considered as references for comparability and reproducibility.
We described each criterion in detail for a clear explanation from three aspects: 
1) what is the criterion? 
2) why do we consider this criterion? 
and 3) how do the current studies meet this criterion? 
This index provides a solution for the problem that there is a lack of generalized criteria to specifically standardize the work in SW technology. 
We conclude that the index is a significantly useful standard to guide and evaluate the work for the normalization of SW technology. 
In future, we will apply this index to guide our other studies in SW technology. 
Finally, we will update this index for new versions by collecting and considering users' feedback.

\bibliography{sample}



\section*{Author contributions statement}

G.L. drafted the major of this manuscript, T. L., X. W., and X. P. contributed to the writing, Z, H. conceived the index and reviewed the manuscript. All authors discussed and analysed the index.

\section*{Competing interests}
The authors declare no competing interests.

\end{document}